# KARB Solution:

# Compliance to Quality by Rule Based Benchmarking

Mohammad Reza Besharati[1]

Mohammad Izadi[2]

1- besharati@ce.sharif.edu, PhD Candidate, Sharif University of Technology, Tehran, Iran, Corresponding Author.

2- izadi@sharif.edu, Assistant Professor, Sharif University of Technology, Tehran, Iran.

## Abstract

Instead of proofs or logical evaluations, compliance assessment could be done by benchmarking. Benchmarks, in their nature, are applied. So a set of benchmarks could shape an applied solution for compliance assessment. In this paper, we introduce the KARB solution: Keeping away compliance Anomalies by Rule-based Benchmarking. By rule-based benchmarking, we mean evaluating the under-compliance-system by its symbolic specification and using a set of symbolic rules (on behalf of semantic logic of evaluation). In order to demonstrate and investigate the manner of KARB solution, we conducted a case study. The IR-QUMA study (Iranian Survey on Quality in Messenger Apps) is defined to evaluate the quality of some messenger apps. The results of evaluations suggest that the Hybrid Method of DD-KARB (with a combination of semantics-awareness and data-drivenness) is more effective than solo-methods and could compute a somehow good estimation for messenger-apps user quality scores. So DD-KARB could be considered as a method for quality benchmarking in this technical context.

## Introduction

Many approaches to compliance assessment have a theoretical nature: they are based on a systematic, rigorous, or formal theory of proof or evaluation (such as Logics, Formal Languages, Proof Systems, Reference Models, and Domain Models). Theories are at a high level of internal-integrity, but they usually suffer from a lack of support and adaptation to the diversity and complexity of real-world cases. In some situations, an aggregation of multiple simpler tools is more successful than a single, rigid, unified, in-depth-designed, and sophisticated one.

In some real-world cases of compliance solving, more lightweight approaches to Formal Specification that support semantic modeling (like generative grammars, production rules, set-theoretic notations, and rewriting logics) could play a key role in overcoming the challenge of semantic diversity and complexity. These approaches support a kind of semantic compilation of several different, diverse, and domain-specific semantic models. For example, by a set of generative rules, we could define the mapping and composition logic of different and independent semantic models.

# About Compliance Checking

Compliance solutions are about assessment, evaluation, verification, validation, or checking of a system, service, process, product, design, organization, or environment regards rules, regulations, laws, standards, specifications, policies, guidelines, protocols, methods, principals, reference-models, etc. [1],[2]. Many application domains benefit and need compliance solutions, including organizations and corporates in these domains: software and IT industry[3], E-Governance[2], finance and banking[4], legal section and professions[5], commerce and trade[6], highly regulated industries (such as food[7] and drug, medical services and devices[8], and construction industry[9]), Complex and Interdisciplinary products and services[10], emerging technology products and services ( such as cyber-physical systems[11], self-driving cars[12], cognitive robotics and agents[13], and smart applications[14]).

There are famous compliance concerns that have been considered by numerous regulations, standards, laws, and acts. Concerns such as security[15], safety[16][17][18], privacy[19][20], data protection[21], accountability[22], responsibility[23], transparency[24], competency[25], anti-piracy[26], anti-corruption[27], antitrust[27], accessibility[28], HCI, quality management and assurance[29][30][31], environmental management[32], sustainability[33], usability[34], human comfort[35], ethics[36], conformance with the disabilities[37], adherence to the children[38] and the elderly[39], simplicity[40] And ease of use are among them.

Modern Paradigms have amplified the necessity for compliance requirements (paradigms such as standardization in business, automation in industries, artificial intelligence and Ubiquitous computing in society, complex systems engineering, socio-technical systems, continuous growth in the economy and social complexity, and quality maturity of services\processes)

In [41], a formal definition (as a 4-tuple) for a special kind of benchmarking is provided. There are formally defined frameworks for compliance checking in legal applications, which are defined in a set-theoretic manner such as some formal systems [5] and Conceptual Modeling of legal texts [5]. Grammar-like and production-rule formalisms have been suggested for automated compliance checking in legal applications [42], [43], [44], [45].

Rules and grammars for architecture conformance checking, especially for the "quality assurance" of the software [46], is another application domain. Some rule-based approaches for architecture selection relate the non-functional requirements, domain requirements, and quality characteristics to architecture styles [47], architecture models [48], architectural patterns [49] and architectural aspects [50].

Circuits and flows are very recurrent modeling approaches in systems engineering. Some researchers consider circuit and flow as a basis for compliance modeling, checking, and benchmarking [51], [52]. There are numerous verification tools and solutions for flow-based models. These tools have served as a means of compliance checking. For example, agent-coordination protocols for crisis situations could be modeled and checked by these tools and solutions [53].

In software engineering, there are some model-based approaches for compliance assurance [54]. These approaches use a modeling notation or framework (such as UML, KAOS[55], and GSN[56]). "Model-based assurance cases" is an approach for safety compliance management. It encompasses a compliance meta-model that contains 'claim' or 'requirement', 'evidence', 'arguments', and 'context' [57], [58]. In [54], the need for a general model of compliance and compliance activity is addressed as a yet open-problem.

In [9] and [59], very close approaches are introduced. The meta-model and system architecture of these approaches are comparable with ours. There are some other meta-models for compliance checking applications and frameworks (for example see [60], [61], [62], [3]).

## About Benchmarking

A benchmark is a common or standard infrastructure to examine, evaluate and compare the reality of solutions, tools or systems by their execution (for some definitions see [63], [64], [65], [66], [67], [68], [69], [70] and [71]. For some examples in other fields see [72], [73], [74], [30],[75] and [76]. Sometimes, simply a measure could be served as a benchmark [77]. In computing The "Procedure", "Measure" and "Computer" are the most common concepts in diverse definitions of benchmarking (Figure 1).

For some early attempts in the history of benchmarking in IT and computing, see [78], [79], [80], [81], [82] references. Also, there was a hot trend for benchmarking in quality assurance, management and continues process improvement [83], [84], which is a progressive journey till now [85], [86], [87].

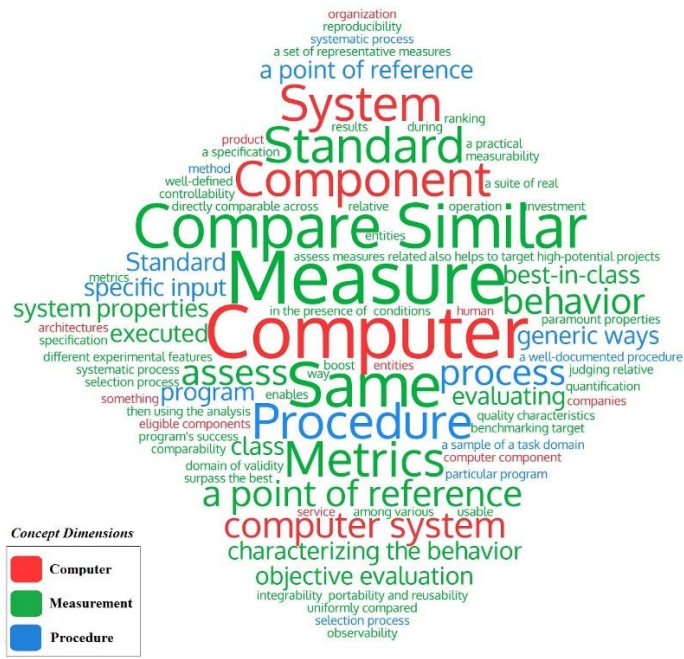

*Figure 1- A segmented word cloud to show the emphasis on different concepts in benchmark definitions.*

From a managerial point of view, benchmarking requires a significant investment in time and perhaps money [84]. So it should be considered a long-term profitable activity and a sort of infrastructure building for a field. For example, in the field of cloud computing, the prior investments in performance measuring tools resulted in already available tools for the new filed (for a case, see [88]); it was a chance but also dedicated attempts began for defining and developing from-scratch benchmarks for cloud computing [89]. Form a managerial point of view, Investment in benchmarking is important [90] and this decision can give benefits to all stakeholders [85].

Successful notions of benchmarking (in each field) have a community that creates, promotes, and uses the benchmarks. Benchmarking could be viewed as an applied manifestation and adoption of communities' knowledge and expertise [91].

There are informal guidelines for ensuring the quality of developed software (for quality attributes such as security, integrity, maintainability, etc.). Formal specification and automatic checking of these guidelines could contribute to better quality assurance of software (see [92]) as an example for security guidelines formalization and automation).

Benchmarks can assess the quality (rather than only functionality) [29], [69], [71], so they are suitable for formal or systematic qualitative analysis of systems. For example, security and compliance benchmarks have been reported [66], [93]. Measuring Productivity of an organization is another example that has qualitative dimensions (such as "the level of customer satisfaction, the quality of the product or the extent to which an organization has the right set of staff" [85]) which could be measured by some systematic approaches [85].

## **Soft Benchmarks**

We could create a knowledge representation once and use it many times (Ontologies are a practical example of this manner of reusability. for example, see [94]). So a community could construct a knowledge representation and serve it as a standard and reusable asset. If this asset helps the community to share their expertise, examine their systems and solutions and evaluate the behavior and other realities of their systems, then we could name it "a soft benchmark". The "soft" part of its name is added to indicate its knowledge-related nature.

Knowledge representations are not limited to ontologies [95]. Formal specifications, such as logical formulations, description logics, semantic networks, and rule-based approaches, are alternatives [96]. Logical models have a share in the approaches of compliance checking. For example, Logical modeling of regulations is a method for rule-representation and checking-automation [97] or using Rules as a paradigm for knowledge representation [98].

A logical theory for a piece of knowledge has the three essential characteristics of a benchmark: 1) it could be a common infrastructure because of the reusable and defined nature of a formal specification, 2) the results of reasoning indicate an examination and evaluation of the system-under-study and provide a basis for comparison between alternative and competing systems, 3) executing reasoning on a logical theory of a piece of knowledge is an execution for the meanings and semantics behind that knowledge. Thinking and mental activities could execute a hypothetical situation. In KARB, rule-based reasoning schema could be viewed as a mimic of these natural procedures (simulation of human auditing by automated and intelligent compliance audit tools is a proper need for compliance industry[99], [66], [2]).

Object models could be served as a semantic model [100], especially also for compliance checking proposes [101], [102]. For example, Fornax objects capture specific rule semantics for compliance checking of building designs [101]. These object models are context, domain, and sometimes system-specific [100]. In the case of Fornax, the objects for hospital design semantics differ from those for airport design [101]. Software Patterns are another representation form or media of technical knowledge. Pattern-based solutions for compliance checking have been addressed in the related researches [103]. Compliance Patterns are a kind of knowledge-capturing tool for compliance assessment.

In KARB, knowledge is represented through an intuitionistic formal semantics, which we call it "semantic logic". Semantic logic was created to capture the semantic and meaning of text [40]. We use it in KARB for knowledge representation. Any knowledge has its semantic and meaning [104], [105], [106], [96]. The knowledge itself is captured, if we capture its semantic and meaning.

Knowledge also has structure [107]. So a meaning structure (or semantic construction) could be a proper candidate for the manifestation of knowledge. Based on and Adopted from [40], we consider semantic and meaning as a construction, lattice[1], or system of realities (=intuitions). Any well-meaning formulation of knowledge represents and references to a combination of entities, things, objects, events, affairs, facts, physics, concepts, cognitions, affections, or any other sort of basic realities and intuitions. So we could aggregately and abstractly consider knowledge as a combination and construction of basic realities and intuitions (with a glue of operators such as logical, structural, modalities, and any needed one). This manner of semantic definition is a constructive and intuitionistic one.

A set of generative rules on basic intuitions (like an axiomatic symbolic logic[2] or a generative grammar) could guide us towards constructing the intended semantic construction of a knowledge representation. In KARB, knowledge representation is made up of two ingredients: formal constructions of symbols (or lattices of symbols) and generative rules. For example, a piece of knowledge is represented in figure 2.

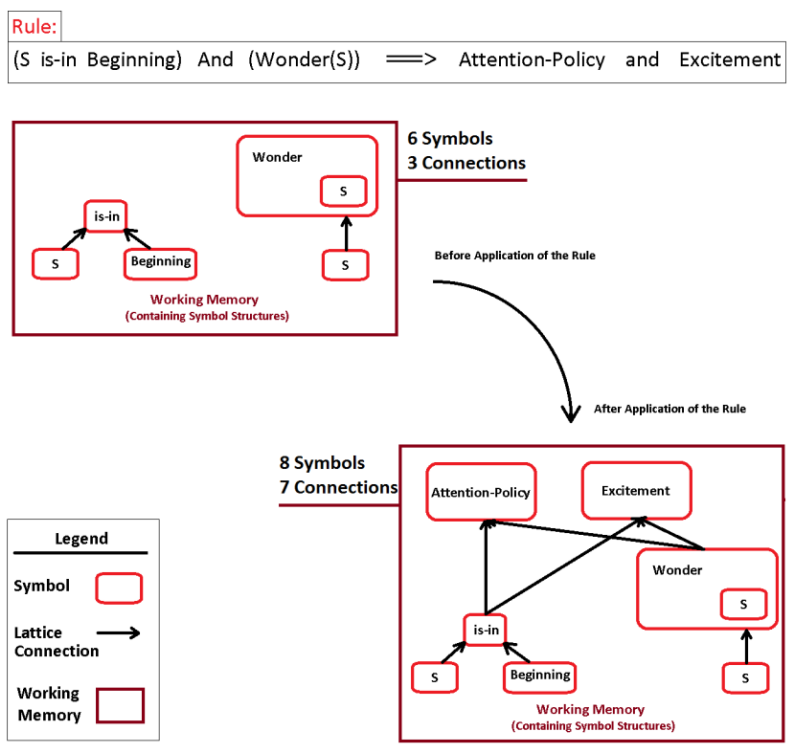

*Figure 2- An example of a semantic logic rule and the effect of its application on working memory. The rule specifies the existence of attention-policy and excitement when a speaker uses a wonderful beginning for a sentence*[40].

---

[1] Lattices are mathematical models for constructions in intuitionistic semantics.

[2] Like a rewriting logic

As an underlying philosophy, in KARB's point of view, the reality and its meanings are composed of statics-related and dynamics-related meanings. Symbolic Constructions capture the statics-related part of knowledge meanings and the generative rules capture the dynamics-related part of knowledge meanings.

## **The KARB solution**

The result of a rule-based benchmarking in KARB is different from other compliance assessment approaches. While other approaches have a result in the form of "yes or no", "correct or incorrect", etc., in KARB we consider a pool of quantities as results (derived and generated symbols). So the overall state of working memory at the end of each benchmarking process indicates the evaluations of the understudy compliance-case. So a rigorous and reasoned evaluation but with diverse dimensions and values would be achieved.

Each aspect of compliance concerns could be addressed by a separate rule-based benchmark. Each benchmark draws and adds a new simple line on the overall picture of compliance assessment scene. A set of multiple, different, and diverse benchmarks could make an applied and realistic compliance assessment of a complex system. This experimental and applied approach to compliance assessment, which relies on rule-based benchmarks, opens a new space for new sorts of innovative, creative, and diverse methods for compliance assessment.

In figure 3, a brief meta-model of KARB is provided. The assessment of each compliance requirement reified by a compliance benchmarking which in turn consists of some concrete rule-based benchmarks. So each compliance requirement declares meanings and semantics for a compliance benchmarking and compliance benchmarking assesses it.

A **Compliance Symbol (CSYM)** abstracts a CCON in a similar sense of atom symbols in Lisp, Objects in OO languages (like Java) and JSON fragments in NoSQL DBs: All of them are units for compositional parts. Each **Compliance Concern (CC)** is in association with some **Compliance Requirements[3] (CR)** which capture the notion and attitude of that concern. For example, Safety is a compliance concern and in a zoo, we could define this Compliance Requirement in accordance with it: the zoo animals must not be able to harm or threat the visitors (see Example 1).

Some **Compliance Rules (CRUL)** aggregately define the operational realization of a **Compliance Requirement**. Each CRU defines a more rigorous, concrete and special obligation than a CR. In KARB, The rules are considered to be finer than requirements. The overall shape of a requirement is made up of the limiting lines of its constituting parts (= rules). Each **Compliance Rule** has some **Compliance Concepts (CCON)** in its definition. In computational mechanisms of KARB, a **Compliance Symbol (CSYM)** abstracts a CCON. By using a glue of operators (logical, structural, modalities and any needed one), we could construct a formal definition of a CRUL from CSYMs of its CCONs (See Example 1).

In KARB, the manner of Formal specification of a CRUL is based on an intuitionistic logic called "semantic logic". Technically, it could be viewed as an axiomatic system on symbols with Brouwer–Heyting–Kolmogorov Interpretation for semantics [108]. Symbols are considered to be on behalf of basic intuitions (concepts, entities, objects, things, events, values, quantities, qualities, etc.) and the under study\ under specification system is viewed as a complex construction of basic intuitions.

---

[3] Compliance Requirement is a known concept in Compliance checking literature and frameworks.

Formally, It is sufficient to consider semantic logic consists of 1) a set of symbols (behalf of basic intuitions), and 2) a set of rules on them. Each rule describes a symbol generation action: when the left-side symbolic structure of the rule is ready in the working memory, the right-side symbolic structure is generated and pushed to the lattice of symbols in the working memory (See Figure 2).

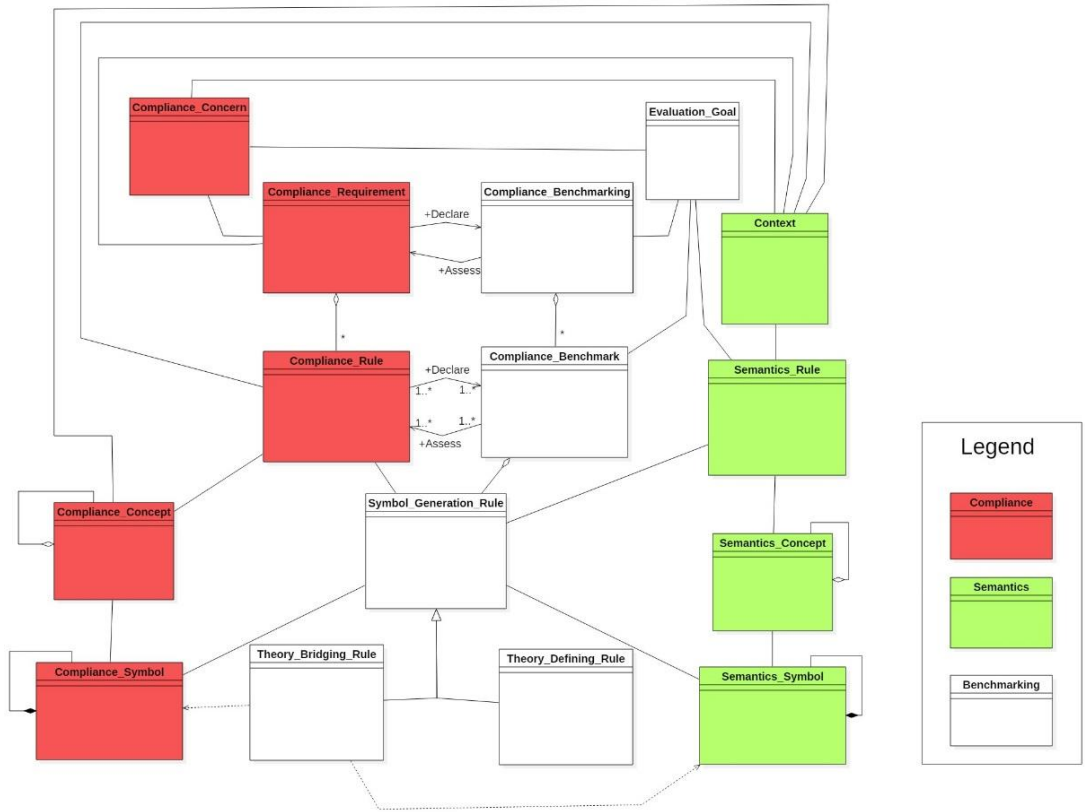

*Figure 3- As a class diagram[4] of elements, the meta-model of KARB solution is depicted. There are three main dimensions for KARB elements: Compliance related (Red), Benchmarking related (White) and Semantics related (Green) concepts.*

**Example 1:**

| The System Under Compliance | a Zoo |
|---|---|
| CC1 | Safety |
| CR1 | The zoo animals must not be able to harm or threat the visitors. |
| CRUL1 | The cage fences must have the proper specification and conditions. |
| CCONs | Cage, fence, proper specification, proper condition |
| CSYMs | CE, FE, PS, PC |
| Formal Specification of CRUL1 | X [IS-A] FE(CE) ==> O[PS(X)] [AND] O[PC(X)] |

---

[4] UML Class Diagram is a popular tool for concept modeling and meta-modeling.

# Case Study: Software Quality Evaluation

Since 1990s, there were a variety of approaches for defining measurable quality: the Quality Function Deployment, the Goal Question Metric, and the Software Quality Metrics [109]. These methods try to shape a general and common framework for quality measurement concerns. But some quality factors are contextual and user dependent[110]. For example, there has been previous research on measuring quality attributes of messenger apps and services from user's perspective (for example see [111], [112], [113], and [114]) or user's behavior (for example see [115], [116] and [117]).

Quality Definitions could be seen as a hierarchical formal system of interrelated concepts [118] or attributes [119] and [120]). This view let us create an explicitly defined conceptual construction for Qualities: a concept-quantized definition for qualities. So a quantification of qualities (which is a well-known but poorly-achieved goal for rigorous software engineering [121]) helps to measure and understand the true level of qualities in each software.

After definition, it is the time for operationalization. Each theoretical concept has its real instances on the ground. User's feedbacks, comments, experiences, requests, requirements, desires, cognitions and intuitions could help this grounding operationalization. So a good quality-definition theory needs a good quality-grounding theory.

There is a semantic gap between definition theories and grounding theories. The first has a neat nature and the last has a scruffy one. How we could bridge the Neats and Scruffies[96]? A glue model could come forward to resolve this challenge. This model must contain the main conceptual elements of the both sides and try to relate them in a gradient conceptual spectrum. Because this is exactly the manner of KARB solution, we could use it as a method to design a software quality evaluation technique. It is also a kind of evaluation for KARB: it demonstrates its usefulness for a real concern or problem of software engineering community.

**Example 2:** as an example, we provide a semantic logic to model the semantics of this scenario:

Using SMS-Based Dynamic Passwords for E-Banking Transactions. This logic contains rules and intuitions from four context theories: Mobile Apps, Deontic Predicate Logic, Security and The System (see figure-4). The stateless model-checking of this scenario semantics (by symbolic-value generations) yields a "false" value. So there is a contradiction in this scenario. Explainable results are provided in the proof construction lattice in figure-5. Although there are some not-mentioned reasoning operations, they are omitted for the sake of simplicity in this preliminary example.

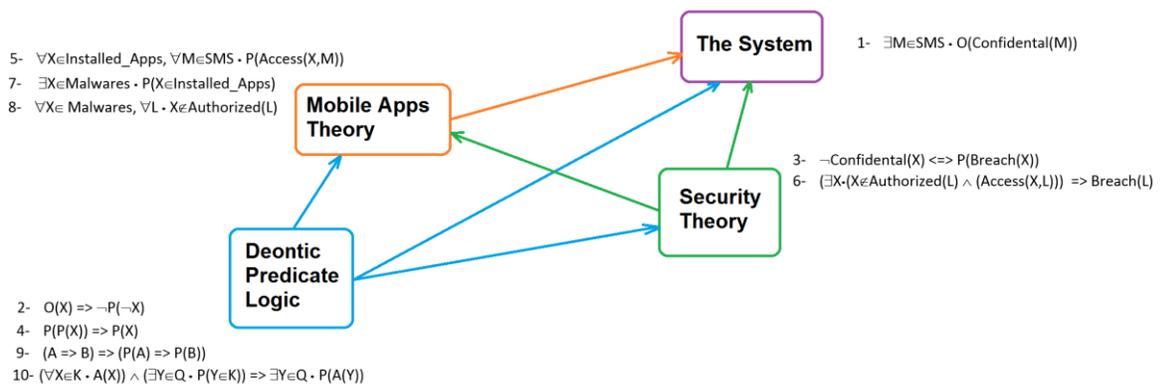

1- ∃M∈SMS • O(Confidental(M))
2- O(X) => ¬P(¬X)
3- ¬Confidental(X) <=> P(Breach(X))
4- P(P(X)) => P(X)
5- ∀X∈Installed_Apps, ∀M∈SMS • P(Access(X,M))
6- (∃X•(X∉Authorized(L) ∧ (Access(X,L))) => Breach(L)
7- ∃X∈Malwares • P(X∈Installed_Apps)
8- ∀X∈ Malwares, ∀L • X∉Authorized(L)
9- (A => B) => (P(A) => P(B))
10- (∀X∈K • A(X)) ∧ (∃Y∈Q • P(Y∈K)) => ∃Y∈Q • P(A(Y))

*Figure 4- The involving semantic theories and the semantic logic for example 2.*

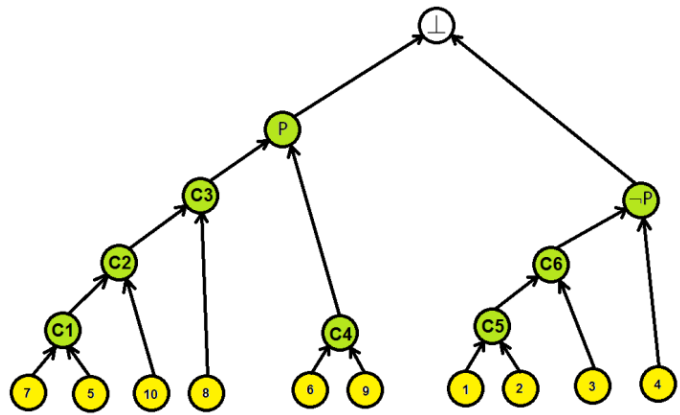

*Figure 5-Proof Construction Lattice for false value from the scenario semantics.*

## DD-KARB

In order to boost the pragmatics of the model, we consider a methodic extension to the principal model of KARB. **D**ata-**D**riven KARB (DD-KARB) incorporates data-calculated weights and values to parametrize the rules. Here we show an example of a parametrized rule.

**Example 3:**

**Rule:**

$$\text{alpha} * (A => B) \quad => \quad \text{beta} * (P(A) => P(B))$$

**Interpretation:**
If alpha instances of (A => B) are generated in the semantic solving of the system (or the weight of (A => B) is equal to alpha), then beta instances of (P(A) => P(B)) must be generated by applying this rule.

Based on expert scores, data-examinations, data-schema, AI pre-trained models and other sources of data-driven models, we could parametrize, annotate and rich our semantics based KARB models with data-driven aspect. The full-fledged methods (by combining the data and semantics aspects) could be better than solo-methods. Each system-context or domain-of-application has its own semantics and data. By DD-KARB, we could record and adapt the both data and semantics aspects of them. This manner of description or declaration, help us to define a hybrid-semantic-core for compliance checking and solving. A hybrid semantics could help us to approach some hard-to-check compliance-requirements by automatic compliance checking solution.

## IR-QUMA Study

Case study is a popular evaluation method in software engineering research. Case studies are frequently used in papers to demonstrate the capabilities of new techniques and methods [122]. In order to demonstrate and investigate the manner of KARB solution, we conducted a case study. The IR-QUMA study (Iranian Survey on Quality in Messenger Apps) is defined to evaluate the quality of some messenger apps. It consists of these stages:

1. **Selection of some Messenger apps**. The selected apps were Telegram, WhatsApp, Eita, Soroush, Bale and some other popular mobile messenger apps in Iranian Cyberspace. We chose these apps for IR-QUMA case study because we had access to a large community of their users.
2. **Data Gathering:** We design an online questionnaire to collect the opinion of users and as a trace or specification of their user-experience. The 7 main questions were about "Absolute Quality", "Relative Quality", "User Satisfaction", "Error-Freeness", "Perceived UI Complexity", "Rationality of Routines", and "Accordance and Usability". Each answer to each question had a range from 1 to 5, to represents choices from "Very Weak" to "Excellent". A depiction of running-average series of users' answer (for a portion of dataset) is depicted in figure-6.
3. **Application of KARB solution**
    a. **Elicitation of involving semantic theories**
    b. **Specification of involving semantic theories.** KARB-based specifications was developed for each of involving semantic theories. A detailed map of involving semantic theories is provided in figure-7. The emphasis was on these theories:

        i. KARB-based Specification of Messenger Apps
       ii. KARB-based Specification of some Quality Terms
     iii. KARB-based Specification of User Behavior
     iv. KARB-based Specification of some pieces of HCI knowledge
      v. KARB-based Specification of Risks and Threats
     vi. KARB-based Specification of Software Platform and Mechanisms
    vii. KARB-based Specification of Cognitive Aspects
   viii. KARB-based Specification of Social Aspects

   c. **Computation and Model Checking:** By KARB solution, we computed some of the compliance anomalies.

4. **Evaluation of the results.** We compared the results with three bases: the experts' judgments, IT reports, and users' opinions.

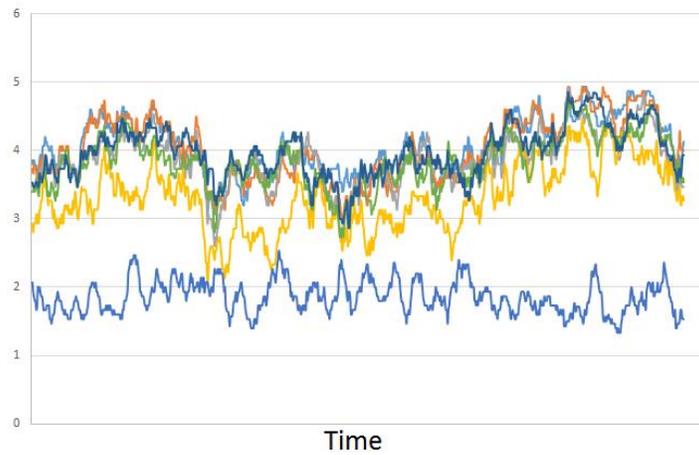

*Figure 6- The running-average series of answers for 7 different questions in the questionaire.*

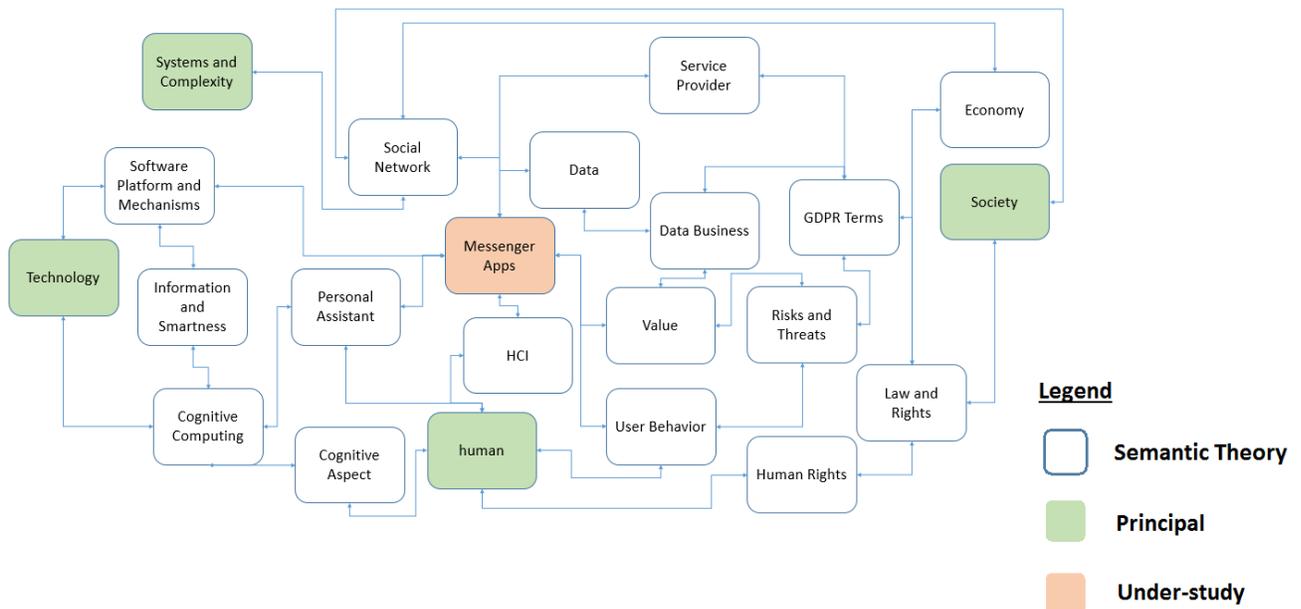

*Figure 7- A detailed map of involving semantic theories, in the IR-QUMA study.*

IR-QUMA study details will be published in its separate report. But in this paper, we used its gathered data and semantic model to conduct some experimentation on deferent quality benchmarks, especially the KARB solution.

## IR-QUMA Data Gathering

A questionnaire was designed to evaluate some quality-related measures, metrics and features from users' experience point of view. The questionnaire was published in popular channels of Iranian mobile social networks, in 10 different messengers (including Telegram, Whatsapp, Instagram, Eita, Soroush, Bale, Gap, IGap, Shaad and Rubika). More than 40 communities of users in these 10 messengers (which are shaping more than 350 micro-communities based on visiting-hour-and-time and spatial partitions) have contributed to this research questionnaire. Total of data is exceeds the level of 7k filled online forms (from more than 7k distinct participants). In our dataset [123], for the sake of data privacy and protection reasons, we hash the name of these messengers randomly by assigning ID-codes from M1 to M10.

A set of statistical analysis, time series analysis, frequency analysis, cluster analysis, classification analysis, geometry locus of data points and topological data analysis are done on these users' opinion data to obtain useful insights. As an analysis example, we sort the data in a temporal order (that conserves the segregation of micro-communities). Then we apply a running average method (with a window-size = 20). So we obtain 7k data-points from different segments of these 350 micro-communities. Each micro-community with its segment-averages has its own footprint in the total space of data-points. Also and again, each messenger app has its own footprint in the total space of data-points.

For example, a correlation locus analysis for two of quality measures for 7 different Messenger apps, in 7k data segments of 350 micro-communities, are depicted in figure-8. Each point is in accordance to measure values obtained from one segment of a micro-community. The blue points are for the mentioned understudy messenger and the red points are for the entire data space (for all understudy messengers). Each axis is demonstrated a 5-level measure value (obtained from averaging of users' opinions in one data segment).

In figure-9, we depict this analysis for two other measures: correctness versus quality. By correctness we mean the error freeness and bug freeness of the messenger app. The results suggest that there is a buffer between "correctness increase\decrease" and "overall quality increase\decrease". This means that the other factors (than correctness) could play an important role in the overall quality of a software.

The histogram of score-instances of user quality judgments for 10 understudy apps are depicted in figure-10. The topological analysis of these 10-curves indicates 6 different curve-clusters, based on change-trends during the 5-levels.

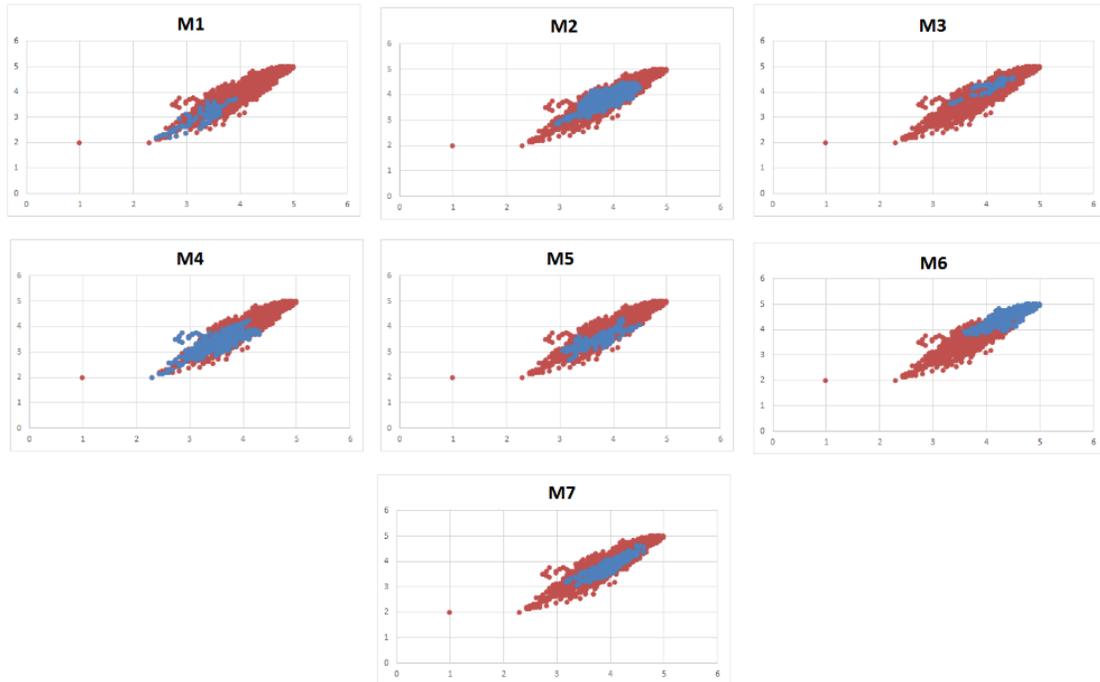

*Figure 8- Correlation between "Absolut quality" measure and "relative quality" measure, from users' point of view, for some messenger apps, from IR-QUMA Data. Each point represents average values of one data segment.*

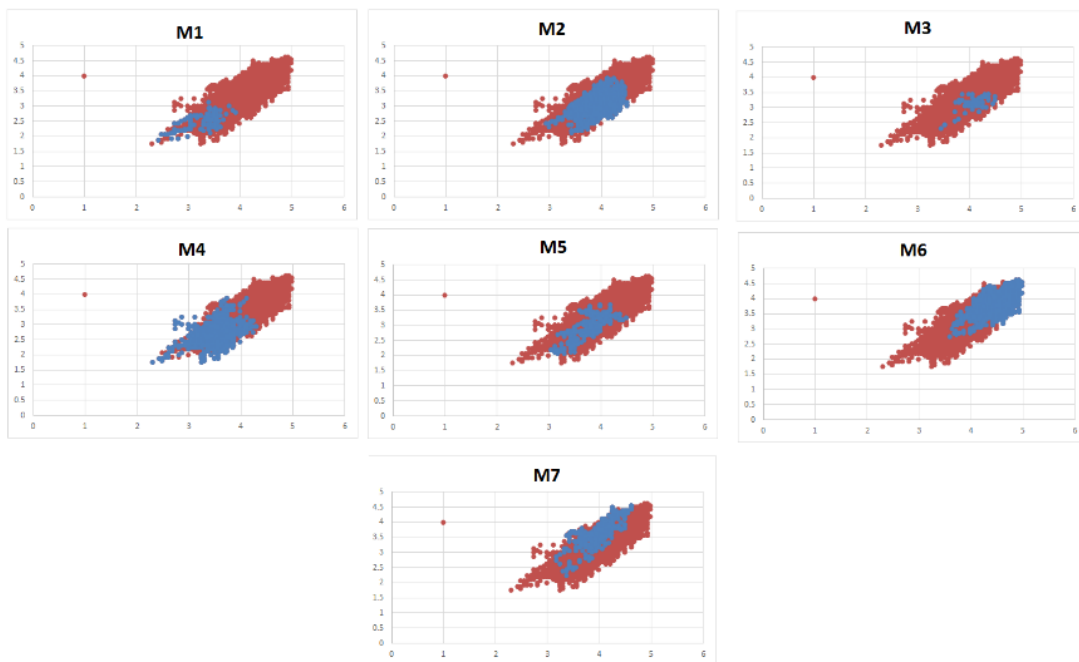

*Figure 9- Correlation between "Correctness" measure and "Absolute Quality" measure, from users' point of view, for some messenger apps, from IR-QUMA Data. Each point represents average values of one data segment.*

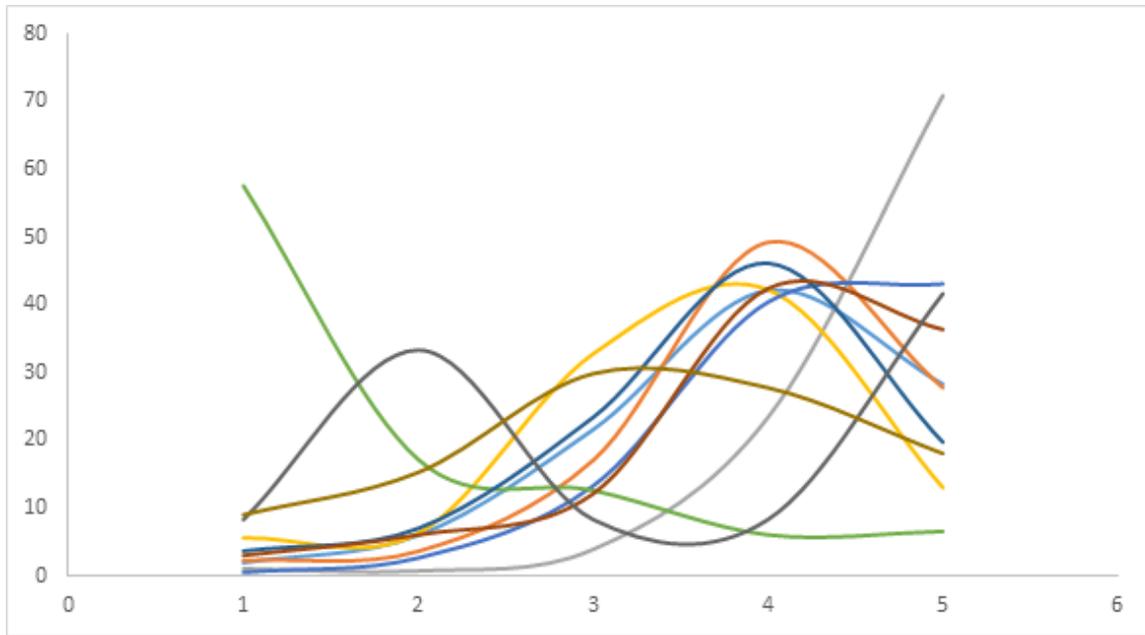

*Figure 10- Histogram Analysis of quality score levels (obtained from users judgments) about "absolute quality" measure, for 10 different understudy messenger apps. Each messenger app has 5 data-points for numeration of 1- very poor, 2-poor, 3-moderate, 4- good and 5-excelent quality scores. These scores reflect the users' experience point of view to the quality of the understudy messenger.*

## Evaluation of the Method

The KARB solution is applied (based on the semantic logic which is specified in previous steps of study) and the performance of it to correctly calculate (or mimic) the user opinions is evaluated (in terms of error percentage of Benchmark-Computed quality scores and user opinions about quality score). We consider this experimentation setting: 4 different methods for Quality Benchmarking and 5 different experiment-runs (for 5 different messengers).

Each user opinion records involves two sections: 1) user opinion about quality score (which is we named the absolute quality score), and 2) quality context. With quality context we means the factors that could affect or relate to user opinion about quality score: age, gender and other information gathered from the user about her\his experience with the messenger (including score of bug-freeness and error-freeness, score of perceived UI complexity, score of rationality of routines, score of usability and …). The value options for all scores in the questionnaires are defined in a 5-valued Likert scale [124] (Likert scale has been used in various domains of software engineering, for example see [125]). The structure of a user opinion record is provided in figure-11.

| Messenger App Name | Absolute Quality Score | Error-freeness Score | Perceived UI Complexity Score | Rationality of Routines Score | Usability Score | Gender and Age |

Figure 11- The structure of User Opinion Record

**Definition 1.** $N_i = $ number of user oponion records about $App_i$

**Definition 2.** $Error\_Percentage(App_i, Method_j) = $

$$\frac{\sum_{k=1}^{N_i} |Benchmak\_Computed\_Quality(App_i, Method_j, Context_k) - User\_Quality\_Oponion_k|}{N_i}$$

The results of evaluation (by above mentioned experimentation setting) is provided in table-1. These results suggest that the data-driven KARB method could reduce the error percentage, significantly. The error reduction curves (for 5 experiments) are depicted in figure-12. The average-curve for these 5 curves show a pseudo-Sigmoid form. This suggests that the Hybrid Method of DD-KARB (with combination of semantics-awareness and data-drivenness) is more effective than solo-methods and could compute a somehow good estimation for messenger-apps user quality scores. So DD-KARB could be considered as a method for quality benchmarking in this technical context.

**Table-1:** The results of evaluation (for Experimentation-Plan-ID-1)

| Experimentation-Plan-ID= 1 | | | Error Percentage | | | | | |
|---|---|---|---|---|---|---|---|---|
| | | | Exp. 1 Telegram | Exp. 2 Eita | Exp. 3 Whatsapp | Exp. 4 Soroush | Exp. 5 Bale | Average |
| Type of Benchmark | Quantitative, Based on Stats | **IR-QUMA (Simple Average)** | 18.4 | 24.3 | 26.3 | 31.8 | 30.1 | 26.2 |
| | Qualitative, Based on Sense | **Expert Quality Scoring** | 17.5 | 23.1 | 25.4 | 30.2 | 29.8 | 25.2 |
| | Declarative, Based on Analysis | **KARB + Hill Climbing** | 11.8 | 15.6 | 15.8 | 17.9 | 16.4 | 15.5 |
| | Hybrid, Based on Solving | **Data-Driven KARB** | 8.8 | 13.2 | 13.5 | 16.8 | 14.0 | 13.3 |

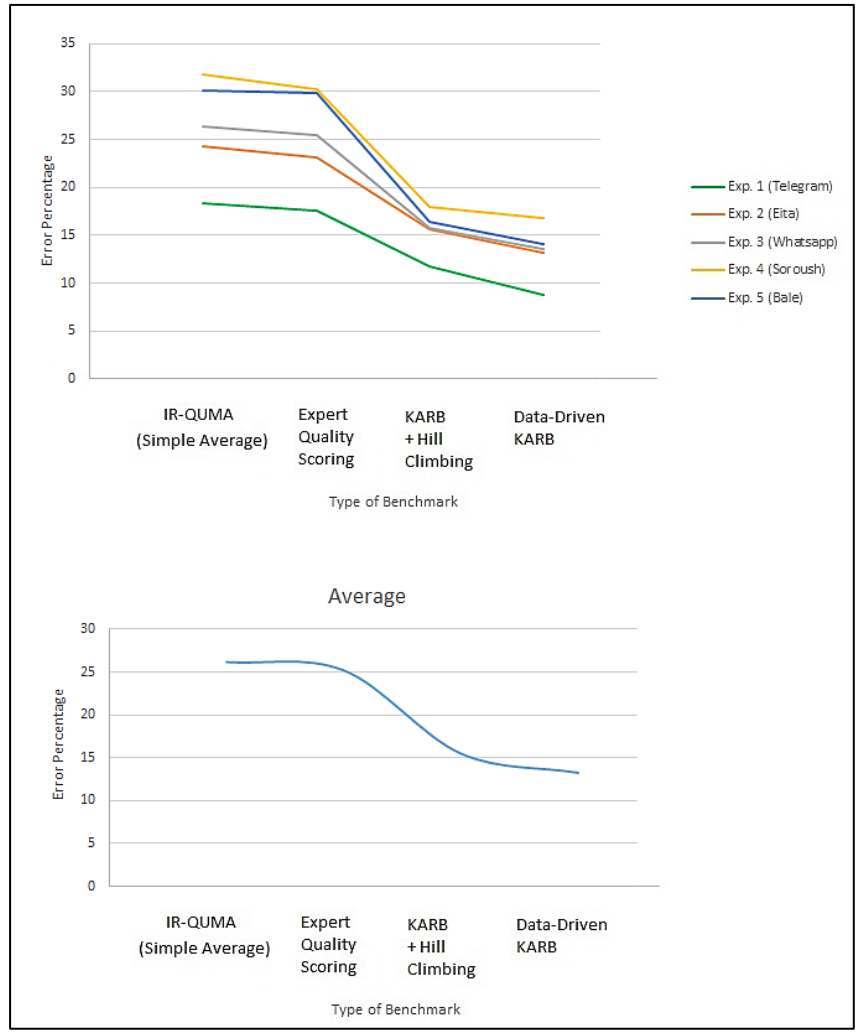

*Figure 12- Error Reduction Curves for 5 different experiments and Average of them.*

# Discussion and Conclusion

The first benchmark uses IR-QUMA context data simple-averages as an estimator for Quality Score. The second benchmark is based on expert judgments about the quality of messengers. The third benchmark is based on a firstly voided-filled DD-KARB rules that take their weight-values from a Hill Climbing optimization algorithm for reaching a local minima of error. The last benchmark, but the first in precision, is based on DD-KARB rules that take their weight-values from two sources: 1- IR-QUMA data values, 2- a lightweight state-space-checking procedure to find somehow good fitting-parameters for (IR-QUMA Data-set, DD-KARB Rule-set). A depiction of some results of this fitting-procedure is shown in figure-13.

So the last benchmark is incorporating these "Method-Features" in a hybrid manner: 1- Semantics-awareness (by KARB), 2- Data-Drivenness (by DD part of DD-KARB and IR-QUMA data), 3- Fitting Solving (by finding fitting-parameters with a lightweight state-space-checking procedure).

The intended semantic landscape of this problem (i.e. Quality Measurement of Messenger apps) is involving more than 10 semantic theories. Without the semantic-framing of the problem, which KARB solution provides it, we couldn't focus on the most relevant parts of the wide semantic landscape of this complex problem. Without using KARB rules which have the role of a kind of declarative-dimensioning in this problem, data-driven solving procedures, verification and model-checking methods couldn't escape from the "state space explosion" [126] in this landscape.

But we escaped with the help of KARB in DD-KARB: a 3-minutes time-consuming process on a conventional PC computer (windows + java + Intel Core i7 Processor) was successfully able to solving a somehow good fitting of a 10K-order complexity-space (IR-QUMA) to a 10G-order value-space of weight-values in DD-KARB rules of this experiment.

We conclude that the hybrid nature of DD-KARB method (in KARB Solution) could help us to solve some complex compliance problems with a lightweight manner and somehow good results (in terms of a low-error compliance-level quality-estimator).

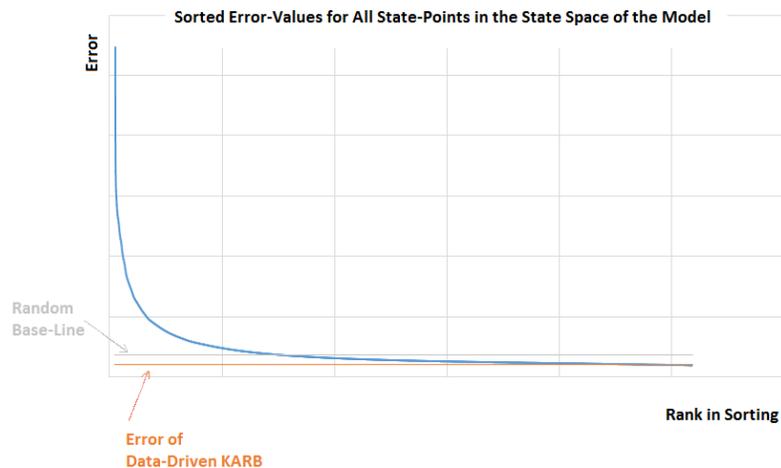

*Figure 13- Depiction of optimization performance of the fitting-solving--procedure in the execution of DD-KARB benchmark in Experiment-1 (Telegram). Each point represents the error of one state in the total state space. The fitting-solving-procedure outperforms the random base-line and finds a state near the exhaustive minima.*